\documentclass[prd,superscriptaddress,amsfonts,amssymb,amsmath,showpacs,twocolumn,showkeys]{revtex4-1}
\usepackage{amsmath}
\usepackage{epsfig}
\usepackage{graphics}
\usepackage[colorinlistoftodos]{todonotes}
\usepackage{color}
\usepackage{graphicx}
\usepackage[font={footnotesize,it}]{caption}
\usepackage[colorlinks=true,
            linkcolor=black,
            urlcolor=blue,
            citecolor=blue]{hyperref}
    
\def\be{\begin{equation}} 
\def\ee{\end{equation}}
\def\bea{\begin{eqnarray}}
\def\eea{\end{eqnarray}}

\def\m{\mu}

\begin{document}

\title{Conformally symmetric traversable wormholes in $f(R,T)$ gravity}

\author{Ayan Banerjee}
\email{ayan\_7575@yahoo.co.in}
\affiliation{Astrophysics and Cosmology Research Unit, University of KwaZulu Natal, Private Bag X54001, Durban 4000,
South Africa.}

   \author{Ksh. Newton Singh}
\email{ntnphy@gmail.com}
\affiliation{National Defence Academy, Khadakwasla, Pune 411023, India,}
\affiliation{Department of Mathematics, Jadavpur University, Kolkata 700032, India.}

\author{M. K. Jasim}
\email{mahmoodkhalid@unizwa.edu.om}
\affiliation {Department of Mathematical and Physical Sciences, University of Nizwa, Nizwa, Sultanate of Oman.}

\author{Farook Rahaman}
\email{rahaman@iucaa.ernet.in}
\affiliation {Department of Mathematics, Jadavpur University, Kolkata 700032, India.}

\begin{abstract}
To find more deliberate $f(R, T)$ astrophysical solutions, we proceed by studying  wormhole geometries under the assumption of spherical symmetry and the existence of a conformal Killing symmetry to attain the  more acceptable astrophysical results. To do this, we consider a more plausible and simple model $f(R,T)=R+2\chi T$, where $R$ is the Ricci scalar and $T= -\rho+p_r+2p_t$ denotes the trace of the energy-momentum tensor of the matter content. We explore and analyze two cases separately. In the first part, wormhole solutions are constructed for the matter sources with isotropic pressure. However, the obtained solution does not satisfy the required wormhole conditions. In the second part, we introduce an EoS relating to pressure (radial and lateral) and density. We constrain the models with phantom energy EoS  i.e. $\omega= p_r/ \rho < -1$, consequently violating the null energy condition. Next, we analyze the model via  $p_t= n p_r $. Several physical properties and characteristics of  these solutions are investigated which are consistent with previous references about wormholes. We mainly focus on energy conditions (NEC, WEC and SEC) and consequently for supporting the respective wormhole geometries in details. In both cases
it is found that the energy density is positive as seen by any static observer. To support the theoretical results, we also plotted several figures for different parameter values 
of the model that helps us to confirm the predictions. Finally, the volume integral quantifier, which provides useful information about the total amount of exotic matter required to maintain a traversable wormhole is discussed briefly.

\end{abstract}


\maketitle

\section{Introduction}\label{intro}
Traversable Lorentzian wormholes are hypothetical tunnels in space-time that connects two regions of the same or disjointed universes. These problems can be attributed in classical general relativityو in which observers may freely traverse. Since wormhole has a long history, but it was developed mainly with the seminal paper by Morris and Thorne \cite{Morris}, in 1988 as a toy model allowing for interstellar travel. In particular, these geometries have a minimal surface area linked to satisfy flare-out condition, which is called throat of the wormhole.  A stress-energy tensor that violates the null energy condition is involving to grip such a wormhole open \cite{Visser}.  Roughly speaking, the matter that violates the weak/null energy conditions called `exotic matter'. Such strange objects exists both in the static \cite{Anabalon:2012tu,Balakin:2010ar,Jamil:2009vn,Cataldo:2002jw} and dynamic \cite{Dehghani:2009xu,Bochicchio:2010df,DeBenedictis:2008qm,GonzalezDiaz:2003pb,Cataldo:2011zn,Hansen:2009kn} cases, and sustained by a single fluid component. The violation of the energy conditions have been supported by many arguments like the quantum field theories such as the Casimir effect, Hawking evaporation and scalar-tensor theories. Though the usage of exotic matter is a problematic issue. Visser \textit{et al.} \cite{Visser:2003yf} have proposed `volume integral quantifier' that how to quantify the total average null energy conditions for wormhole maintenance. 
    
However, a static wormhole without violating the energy conditions in the framework of Einstein General Relativity is still an open problem, which can be motivated to minimize the usage of exotic matter by applying the cut and paste technique, which was proposed by Visser \cite{Visser:1989kh,Visser:1989kg}. The proposal was to restrict the exotic fluid at the wormhole throat. There were another solution came from Kuhfittig \cite{Kuhfittig:2002ur,Kuhfittig:1999ur} to hamper the exotic fluid of an arbitrary thin region by imposing a condition on $b'(r)$ to be close to one at the throat.    
    
One may also follow a more conventional method to address the issue in an alternatives theories of gravity. The physical incentives for these amendments of gravity are based on gravitational actions which are linked to the possibility of a more realistic illustration of the gravitational fields near curvature singularities. The main purpose of this approach lies on the assumption that matter threading the wormhole satisfies the energy conditions.   Due to the effective stress-energy tensor, the field equations have to rewritten in a form that represented as a sum of the standard fluid plus the new terms coming from the modified theory. In this context, several wormhole solutions were analyzed in various modified gravity theories such as  $f(R)$ gravity \cite{Lobo:2009ip,DeBenedictis:2012qz,Mazharimousavi:2012xv,DiCriscienzo:2013ria,Eiroa:2015hrt,Pavlovic:2014gba},  $f(R)$ gravity wormhole with noncommutative geometry \cite{Jamil:2013tva}, $f(T)$ gravity \cite{Sharif:2013exa,Sharif:2014bsa,Sharif:2013lya},  noncommutative geometry \cite{Rahaman:2012pg,Zubair:2019qqz,bhar2014}, Lovelock solutions \cite{Dehghani:2009zza,Matulich:2011ct,Zangeneh:2015jda,Mehdizadeh:2016nna} and in others. 

In this article, we are particularly interested in $f(R,T)$ gravity \cite{Harko:2011kv}, where the Lagrangian is an arbitrary function of Ricci scalar $R$ and the trace of the energy-momentum tensor $T$. This theory has been tested from cosmology to astrophysics and are more manageable compared to $f(R)$ theories. Recently, this model has been extensively investigated, such as thermodynamics
properties~\cite{frttd1,frttd2,frttd3}, energy
conditions~\cite{frtec1,frtec2,frtec3}, cosmological solutions
based on a homogeneous and isotropic space--time through a
phase--space analysis~\cite{phsp},  a cosmological solution via a
reconstruction program~\cite{frtrec1,frtrec2}, anisotropic
cosmology~\cite{frtani2,frtani3}, a cosmological
solution via an auxiliary scalar field~\cite{frtaf}, the study of
scalar perturbations~\cite{frtsp}.  But, a serious shortcoming in this modification has been the non-conservation of the energy-momentum tensor. Non-conservation of the energy-momentum tensor is also found in relativistic diffusion models (Ref. \cite{Calogero} and references therein).  This fact has to be stressed
because it demonstrates somehow a limitation for this class of theories. However, consistent cosmological solutions are in favor of this theory.  For detailed review of $f(R,T)$- gravity one may refer to \cite{Houndjo:2011fb}. In the following, a static wormhole solution have been obtained by Moraes \& Sahoo \cite{Moraes:2017mir}. Also, a charged wormholes in $f(R,T)$ gravity has been proposed recently in \cite{Moraes:2017rrv,Banerjee:2020uyi}.

The theoretical construction of wormhole geometries lies on the fact that one has a desired metric, which have to solve by fixing the form of the metric potential functions or by using a precise equation of state that relates the pressure with the energy density, and then solve Einstein’s field equations. In our work an exact solutions by assuming spherical symmetry and the existence of a \textit{non-static} conformal symmetry have been studied in an alignment of systematic approach that was  considered previously by Boehmer \textit{et al} \cite{Boehmer:2007rm,Boehmer:2007md}. The study of conformal symmetry gives a natural link between geometry and matter through the Einstein field equations. It is for this reason the vector $\xi$ has been specified as the generator of  this conformal symmetry, then the metric $g$ is conformally mapped onto itself along $\xi$, which is interpreted into the following relationship
\begin{equation}\label{eq1}
\mathcal{L}_{\xi}g_{ij}= \psi g_{ij},
\end{equation}
where $\mathcal{L}$ is the Lie derivative operator of the metric tensor and
$\psi$ is the conformal killing vector. Also, for a static metric, we have noted that neither $\xi$
nor $\psi$ need to be static. This approach was used in \cite{Herrera1984,Herrera1985}, to show that for a one-parameter group of conformal motions, the EoS is uniquely determined by the Einstein equations. Later, this particular exact solution was extended  by Maartens \&  Maharaj \cite{Maartens}, for static spheres of charged imperfect fluids with assuming space-time admits a conformal symmetry. Very recently, Kuhfittig \cite{Kuhfittig:2016pyx,Kuhfittig:2015cea} have
studied wormholes admitting a one-parameter group of conformal motions.

The plan of this paper is as follows: After the introduction in Section \ref{intro}, we briefly 
review the field equations of $f(R,T)$ gravity, in particular when the matter is minimally coupled to the curvature in a specific form, are presented in Section \ref{sec2}. 
In Section \ref{sec3}, we discuss a specific spacetime metric (Morris-Thorne metric) of a spherically symmetric traversable wormhole and the basic mathematical criteria.
In Section \ref{sec4}, exact general solutions are deduced using static conformal symmetries. In section \ref{sec5}, we present the unique exterior vacuum solution.
Then, we study the wormhole models from different hypothesis for their matter content;  specifically for isotropic pressure and linear EoS relating the energy density and the pressure anisotropy in section \ref{sec6}. Finally, in Section \ref{sec7}, we conclude.\\

\section{Basic mathematical formalism of the $f(R,T)$ Theory} \label{sec2}
In this section, we start by writing the general action for $f(R,T)$
modified gravity in four-dimensional spacetime. The full action is given by Harko \emph{et al}  \cite{Harko:2011kv} (with geometrized units $c = G = 1$)
\begin{equation}\label{eq2}
S=\frac{1}{16\pi}\int d^{4}xf(R,T)\sqrt{-g}+\int
d^{4}x\mathcal{L}_m\sqrt{-g},
\end{equation}
where $f(R,T)$ is an arbitrary function depends on a generic function of $R$ and $T$, the Ricci scalar and the trace of the energy momentum tensor $T_{\mu\nu}$, respectively. From the matter Lagrangian density
$\mathcal{L}_m$, we defined the energy-momentum tensor as follows
\begin{equation}\label{eq3}
T_{\mu\nu}= -\frac{2}{\sqrt{-g}}\frac{\delta\left(\sqrt{-g}\mathcal{L}_m\right)}{\delta g^{\mu\nu}}.
\end{equation}
Following the argument in \cite{Harko:2011kv}, we assume that the Lagrangian density $\mathcal{L}_m$ depends only on the metric components $g_{\mu\nu}$ and not on its derivatives, we obtain
\begin{equation}\label{eq4}
T_{\mu\nu}= g_{\mu\nu}\mathcal{L}_m-2\frac{\partial \mathcal{L}_m }{\partial g^{\mu\nu}}.
\end{equation}
Now, by variation of the action $S$ given in Eq. (\ref{eq2}) with respect to the metric $g_{\mu\nu}$, to obtain the gravitational field equation for $f(R,T)$ gravity as:
\begin{widetext}
\begin{eqnarray}\label{eq5}
f_R (R,T) R_{\mu\nu} - \frac{1}{2} f(R,T) g_{\mu\nu} + (g_{\mu\nu}\Box - \nabla_{\mu} \nabla_{\nu}) f_R (R,T)= 8\pi T_{\mu\nu} - f_{T(R,T)} T_{\mu\nu} - f_{T(R,T)}\Theta_{\mu\nu} ,
\end{eqnarray}
\end{widetext}
where $f_R (R,T)= \partial f(R,T)/\partial R$,
$f_T (R,T)=\partial f(R,T)/\partial T$,
$\Box \equiv \partial_{\mu}(\sqrt{-g} g^{\mu\nu} \partial_{\nu})/\sqrt{-g}$,
$R_{\mu\nu}$ is the Ricci tensor, $\nabla_\mu$ denotes the covariant derivative with respect to the metric
$g_{\mu\nu}$ and $\Theta_{\mu\nu}= g^{\alpha\beta}\delta
T_{\alpha\beta}/\delta g^{\mu\nu}$ .

Performing a covariant divergence of (\ref{eq5}) which yield \cite{Baffou:2013dpa,Singh:2013bpa,Sharif:2014ioa,Baffou:2017pao,Mishra_2018}
\begin{widetext}
\begin{eqnarray}\label{eq6}
\nabla^{\mu}T_{\mu\nu}&=&\frac{f_T(R,T)}{8\pi -f_T(R,T)}[(T_{\mu\nu}+\Theta_{\mu\nu})\nabla^{\mu}\ln f_T(R,T)\nabla^{\mu}\Theta_{\mu\nu}-(1/2)g_{\mu\nu}\nabla^{\mu}T].
\end{eqnarray}
\end{widetext}
For this purpose we assume the matter content of the  wormhole solution is an anisotropic fluid and one can write the energy momentum tensor as
\begin{equation}\label{eq7}
T_{\mu\nu}=(\rho+p_r)u_\mu u_\nu-p_t g_{\mu\nu}+ (p_r-p_t)g_{\mu\nu},
\end{equation}
where $\rho$ is the energy density with $p_r$ and $p_t$ representing the radial and tangential pressures of the fluid, $u^{\mu}$ is the four-velocity such
that $u^{\mu}u_{\mu} = 1$ and $u^\mu\nabla_\nu u_\mu=0$. In this way, one can choose the matter Lagrangian density  as $\mathcal{L}_m = -\mathcal{P}$, where $\mathcal{P} =\frac{1}{3} (p_r+2p_t)$ which is more generic, in the sense that they do not imply the vanishing of the extra force, which yields 
$\Theta_{\mu\nu}=-2T_{\mu\nu}- \mathcal{P}g_{\mu\nu}$.

In the present work, we focus our attention on the simplified and linear functional form of  $f(R,T)=R+2\chi T$, as suggested by Harko \textit{et al} ~\cite{Harko:2011kv}, where $\chi$ is a constant. The chosen form has been broadly applied in many cosmological solutions of $f(R,T)$ gravity~\cite{Houndjo:2011fb}. Our ansatz for the function $f$, the Eq. (\ref{eq5}) becomes \cite{Moraes:2017mir,Moraes:2017rrv}
\begin{equation}\label{eq8}
G_{\mu\nu}=8\pi T_{\mu\nu}+\chi
Tg_{\mu\nu}+2\chi(T_{\mu\nu}+pg_{\mu\nu}),
\end{equation}
where $G_{\mu\nu}$ is the Einstein tensor. If we set $\chi=0$, then one can easily recover the general relativistic result. It is straightforward to see that for the particular choice of $f(R,T)= R+2\chi T$, Eq. (\ref{eq4}) leads to the  form
\begin{equation}\label{eq9}
(8\pi+2\chi)\nabla^{\mu}T_{\mu\nu}=-2\chi\left[\nabla^{\mu}(pg_{\mu\nu})+\frac{1}{2}g_{\mu\nu}\nabla^{\mu}T\right].
\end{equation}
Regarding the Bianchi identity, obviously in $f(R,T)$ gravity, the covariant derivative of the energy-momentum tensor is not null in general.  But substituting $\chi=0$ in Eq. (\ref{eq9}), one can see that the energy-momentum tensor is conserved as in case of general
relativity.

\section{Traversability conditions and general remarks for wormholes} \label{sec3}
The spacetime ansatz for seeking traversable static spherically symmetric wormholes is the Morris-Thorne metric \cite{Morris}, which can be written as
\begin{equation}
ds^2=-e^{\nu(r)}dt^2+\frac{dr^2}{1-\frac{b(r)}{r}}+r^2(d\theta^2+\sin^2\theta
d\phi^2),\label{eq10}
\end{equation}
where $\nu(r)$ and $b(r)$ are the redshift and the shape functions, respectively. The redshift function $\nu(r)$  must be finite everywhere, in order to ensure the absence of horizons and singularities. The essential characteristics of a wormhole is the shape function $b(r)$ which determine the shape of the wormhole must satisfy the condition $b(r = r_0$) = $r_0$ at the throat $r_0$ where $r_0 \leq r \leq \infty$. For the existence of standard wormholes, the shape function should satisfy the ``flaring-out condition", given by
\begin{equation}\label{eq11}
\frac{b(r)-rb^{\prime}(r)}{b^2(r)}>0,
\end{equation}
which reduces to $b^{\prime}(r_0) < 1$ at the throat $r = r_0$. Here  the prime denotes the derivative with respect to the radial coordinate $r$. Moreover, finiteness of the proper radial distance, $\ell(r)$ defined by
\begin{equation}\label{eq12}
\ell (r) = \pm \int^{r}_{r_0}{\frac{dr}{\sqrt{1-\frac{b(r)}{r}}}}  ,
\end{equation}
is required to be finite everywhere. It is important to note that `$\ell$' the proper distance is greater than or equal to the coordinate distance, i.e. $ \mid \ell (r) \mid$ $\geq$ $r-r_0$ where the $\pm$ signs refer to the two asymptotically flat regions which are connected by the wormhole. Since, $\ell\enspace$  decreases from  $\ell = +\infty$ to at the throat of the wormhole $\ell = 0$, and then from $\ell = 0$ to $\ell = -\infty$.

Following the metric Eq. (\ref{eq10}), the Einstein tensor, $G_{\mu\nu}$ = $R_{\mu\nu}-\frac{1}{2}R g_{\mu\nu}$ then reduce to the  following non-zero components
\begin{eqnarray}
G_0^{0}&=& \frac{b^{\prime} (r)}{r^{2}},\label{eq13}\\
G_1^{1}&=& -\frac{b(r)}{r^{3}} +\left(1-\frac{b(r)}{r}\right)\frac{\nu^{\prime}}{r},\label{eq14}\\
G_2^{2}&=& \frac{1}{4}\left(1-\frac{b(r)}{r}\right)\left[\nu^{\prime 2}+2\nu^{\prime\prime}
-2\frac{b^{\prime}r-b}{r^2(r-b)} \right.  \nonumber\\
&& \left.
 -\frac{b^{\prime}r-b}{r(r-b)}\nu^{\prime}+\frac{2\nu^{\prime}}{r}\right],\label{eq15}\\
G_3^{3}&=& G_2^{2},\label{eq16}
\end{eqnarray}
where primes stand for derivation with respect to the radial coordinate $r$.

\section{The Conformal Killing Vector (CKV)}\label{sec4}
Construction of wormhole can be straightforwardly generalised to conformal theories containing matter fields. Based on the assumption that spherically symmetric static space-time possesses a conformal symmetry and identify its essential mathematical structure, one can simplify the treatment of the problem and define its basic mathematical structure \cite{Herrera1984, Maartens}.  The existence of a Killing vector laid constraints on the influences of curvatures of the manifold and symmetry.  If we consider a static metric, the vector fields $\xi$ and $\psi$ are not necessary to be static. So, the Eq. (\ref{eq1}) can be written in a simple way as
\begin{equation}
\mathcal{L}_{\xi} g_{ij} = \xi_{i;j}+ \xi_{j;i} = \psi g_{ij},\label{eq17}
\end{equation}
where the Lie derivative operators $\xi_i = g_{ik}\xi^k$ and  $\mathcal{L}$ describes the interior gravitational field of a wormhole configuration.  Constants of the motion may be determined by the Killing vectors i.e. quantities that will be constant along any given geodesic.  Furthermore, the conformal vectors can be obtained when (i) $\psi=0$, then Eq. (\ref{eq17}) gives the Killing vector, (ii) $\psi=$ constant gives homothetic vector, and (iii) when $\psi=\psi(\textbf{x},t)$ gives conformal vectors.

After introducing conformal Killing vector Eq. (\ref{eq17}) into the metric Eq. (\ref{eq10}), without a loss
of generality  provides the following solutions
\[ \xi^1 \nu^\prime =\psi(r),~~~ \xi^4  = {\rm const.},~~~\xi^1  = \frac{\psi r}{2}, ~~~ \xi^1 \lambda^\prime + 2 \xi^1 _{,1} =\psi(r), \]
where $1$ and $4$ represents the spatial and temporal coordinates
$r$ and $t$, respectively.

These, in turn, imply that
\begin{eqnarray}
e^\nu  &=& C_2^2 r^2, \label{eq18}\\ \left(1-\frac{b(r)}{r}\right)^{-1}  &=&
\left[\frac {C_3} {\psi}\right]^2,  \label{eq19} \\ \xi^i &=& C_1
\delta_4^i + \left[\frac{\psi r}{2}\right]\delta_1^i, \label{eq20}
\end{eqnarray}
where $C_1$, $C_2$ and $C_3$ are constants of integration. Notice that if the Eq. (\ref{eq19}) written in terms of the shape function $b(r)$, then the conformal factor is zero at the throat, i.e. $\psi(r_0) =0$. It should be emphasized that the solutions given by Eqs. (\ref{eq18}) and (\ref{eq19}),  and using the above conformal relations  relating the form and redshift
functions $\nu(r) ={1 \over 2} \ln(C^2 r^2)-\lambda \int{\frac{dr'}{r' \psi(r')}}$ places a strong constraint on
the specific choices of the wormhole geometries. From the above relation, it is obvious that imposing the choices for the
redshift function, one may deduce the form function
and the conformal factor also.

The strong constraints on the wormhole geometry will be imposed by the existence of conformal motions. Consider the above energy-momentum tensor and the Morris-Thorne metric Eq. (\ref{eq10}), the
generalized gravitational field equations  (\ref{eq8}) give the following field equations
\begin{eqnarray}
&& \frac{b^{\prime}}{r^{2}} = \left(8\pi+\chi\right)\rho-\chi\left(p_r+2p_t\right),\label{eq21}\\
&& \left[1-\frac{b}{r}\right]\frac{\nu^{\prime}}{r}-\frac{b(r)}{r^{3}} = \chi \rho+\left(8\pi+3\chi\right)p_r+2\chi p_t,\label{eq22}\\
&& \frac{1}{4}\left[1-\frac{b}{r}\right]\left[\nu^{\prime 2}+2\nu^{\prime\prime}-2\frac{b^{\prime}r-b}{r^2(r-b)}-\frac{b^{\prime}r-b}{r(r-b)}\nu^{\prime} \right.  \nonumber\\
&& \left. +\frac{2\nu^{\prime}}{r}\right] = (\rho+ p_r)\chi +\left(8\pi+4\chi\right) p_t. \label{eq23}
\end{eqnarray}

\begin{figure}[t]
\includegraphics[width=7.5cm]{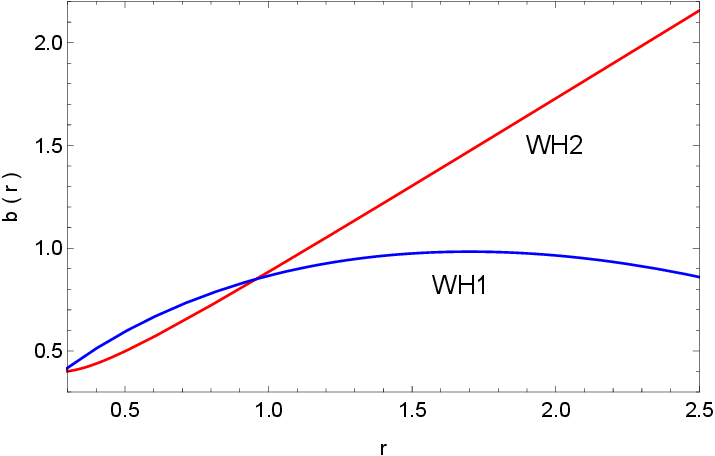}
\caption{Variation of the shape function with $A=1.23$, $\chi =-2,~ \omega =-2, ~c_3=7.74$ (WH1) and $B=-0.44,~\chi =-2,~n=-0.4,~c_3=-10$ (WH2).}\label{fig1}
\end{figure}

\begin{figure}[t]
\includegraphics[width=7.5cm]{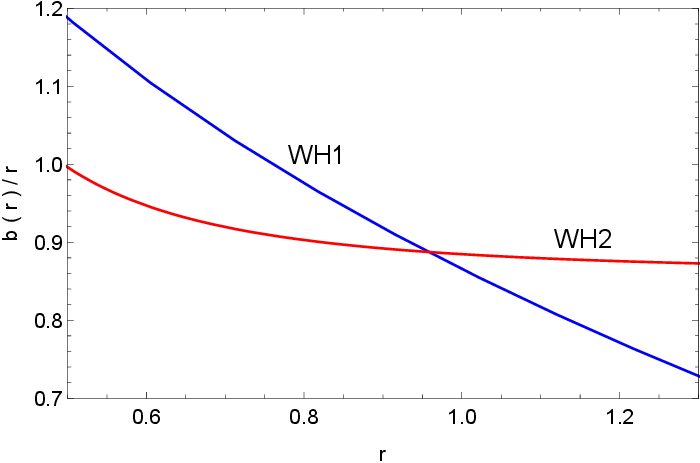}
\caption{Variation of $b(r)/r$ with $A=1.23$, $\chi =-2,~ \omega =-2, ~c_3=7.74$ (WH1) and $B=-0.44,~\chi =-2,~n=-0.4,~c_3=-10$ (WH2).}\label{fig2}
\end{figure}

Thus, using the expression (\ref{eq18})-(\ref{eq20}), in the above Eqs. (\ref{eq21})-(\ref{eq23}), we obtain a set of field equations as follows
\begin{eqnarray}
-\frac{2\psi\psi'}{rC_3^2}-\frac{\psi^2}{r^2C_3^2}+\frac{1}{r^2}
&=& \rho_{\text{eff}},\label{eq24}\\
\frac{3\psi^2}{r^2C_3^2}-\frac{1}{r^2}&=&  p_{\text{eff}},\label{eq25}\\
\frac{\psi^2}{r^2C_3^2}+\frac{2\psi\psi'}{rC_3^2} &=&  \mathcal{P} _{\text{eff}}, \label{eq26}
\end{eqnarray}
where the $\rho_{\text{eff}}$ and $p_{\text{eff}}$ are given by
\begin{eqnarray*}
\rho_{\text{eff}}&=& \left(8\pi+\chi\right)\rho-\chi\left(p_r+2p_t\right),\\
 p_{\text{eff}}&=& \chi \rho+\left(8\pi+3\chi\right)p_r
+2\chi p_t,\\
 \mathcal{P} _{\text{eff}}&=&(\rho+ p_r)\chi +\left(8\pi+4\chi\right) p_t.
\end{eqnarray*}
In addition to other essential characteristics of a wormhole solution, the violation of the null energy condition  (NEC) at the throat of the wormhole is a generic feature. Therefore, such energy conditions are deemed important since they lead to physical requirements on matter.

Considering the $f(R)$ gravity, Garcia and Lobo \cite{MontelongoGarcia:2010xd} showed that nonminimal coupling minimizes the violation of the NEC of normal matter at the throat.
Moreover, Einstein-Cartan theory attracted a good deal of attention in wormhole solution without invoking exotic matter \cite{Bronnikov:2015pha}.  Quantum effects also produce violations of the classical energy conditions, amongst which the popular one is Casimir effect.

In the context of the local energy conditions, we examine the
the violation of NEC, $T^{\text{eff}}_{\mu\nu} k^{\mu}k^{\nu} \geq 0$, where $k^{\mu}$ is any null vector  and $T_{\m\nu}$ is the usual Hilbert stress-energy-momentum tensor. In combination to the above expression we have
\begin{eqnarray}
8 \pi\left(\rho_{\text{eff}}+p_{\text{eff}}\right)=-\frac{2\psi\psi'}{rC_3^2}+\frac{2\psi^2}{r^2C_3^2},
\end{eqnarray}
which evaluated at the throat imposes the following condition ($\psi^2)' >0$.

\begin{figure}[t]
\includegraphics[width=7.5cm]{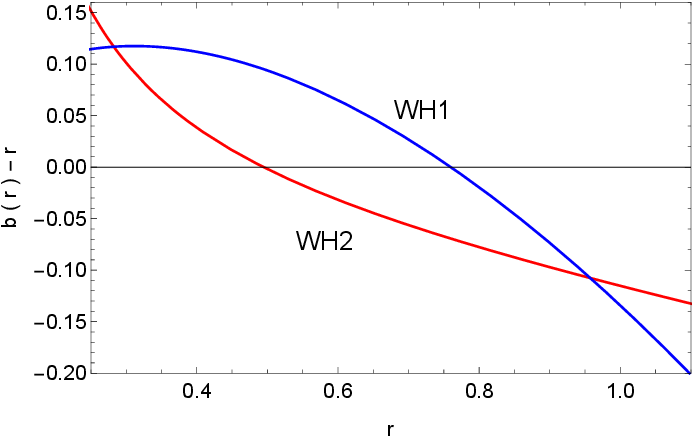}
\caption{Variation of $b(r)-r$ with $A=1.23$, $\chi =-2,~ \omega =-2, ~c_3=7.74$ (WH1) and $B=-0.44,~\chi =-2,~n=-0.4,~c_3=-10$ (WH2).}\label{fig3}
\end{figure}

\begin{figure}[t]
\includegraphics[width=7.5cm]{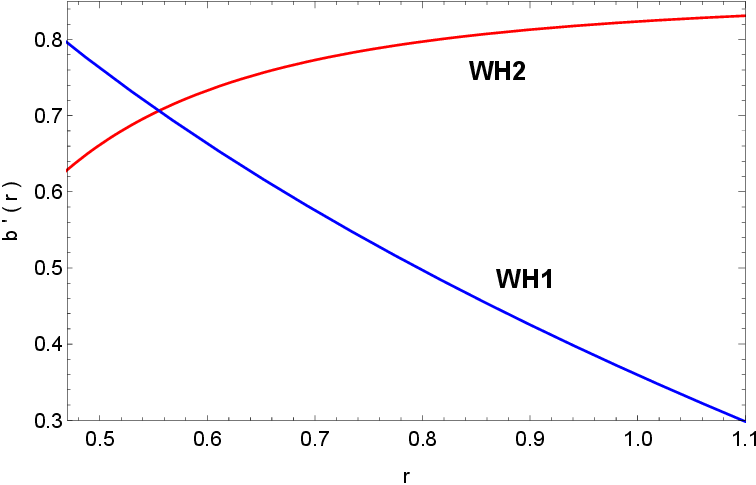}
\caption{Variation of $b'(r)$ with $A=1.23$, $\chi =-2,~ \omega =-2, ~c_3=7.74$ (WH1) and $B=-0.44,~\chi =-2,~n=-0.4,~c_3=-10$ (WH2).}\label{fig4}
\end{figure}

\section{Thin shell around traversable $f(R,T)$ wormhole}\label{sec5}
We shall model specific static wormholes by matching an interior geometry, with an exterior Schwarzschild vacuum solution, at a junction interface $\sum$= $\sum_{+} = \sum_{-}$. Our aim here is to restrict the dimensions of these wormholes not to arbitrarily large. For this, the exterior Schwarzschild is given by
\begin{equation}
ds^2=-\left(1-\frac{2M}{r}\right)dt^2+\frac{dr^2}{\left(1-\frac{2M}{r}\right)}+r^2(d\theta^2+\sin^2\theta
d\phi^2),
\end{equation}
which we shall match with the interior spacetime given in Eq. (\ref{eq10}).

Following the standard junction-condition formalism in (3 +1)-dimensional spacetime \cite{Musgrave:1995ka,Sen:1924,Lanczos:1924,Israel:1967}, one can consider two pseudo-Riemannian manifolds with a radius greater
than the event horizon radius, and paste them at the hypersurface to create a geodesically complete manifold. If such boundaries are identified, then a natural match of manifolds can be done, with two regions connected by a throat of radius, where the exotic matter is located \cite{Poisson:1995sv,Bejarano:2006uj,Eiroa:2003wp,Lemos:2008aj,Rahaman:2008xs,Forghani:2018gza,Banerjee:2016blr,Banerjee:2012aja}. 
Beyond GR, the junction formalism requires to be generalized and several conditions that should be fulfilled for the specific theory of gravity under consideration. For example in  $f(R)$ gravity, the junction conditions tend not to always coincide with those  of general relativity \cite{Deruelle:2007pt,Senovilla:2013vra,Goswami:2014lxa} (see also Refs. \cite{Velay-Vitow:2017odc} for  $f(T)$ gravity).

To understand the above in some details we would like to point out a special feature of the thin-shell structure. More tactically for a geodesically complete  thin-shell wormholes, the Riemann tensor is divergent at the thin-shell where the throat is
located \cite{Poisson}. To see this let $\Sigma$ be a non-null hypersurface layer, and suppose the coordinate system on both sides of the hypersurface to be the same then $\mathcal{X}^\mu_{\pm}$  defines the jump of a quantity $Z$ as
\begin{equation}
  [Z]= Z (\mathcal{X}^{+}_\mu) |_{\Sigma}-Z (\mathcal{X}^{-}_\mu) |_{\Sigma}.
\end{equation}
Then, the distribution of matter reads
\begin{equation}
  T_{\mu\nu}= \Theta(x)T^{+}_{\mu\nu} +\Theta(-x)T^{-}_{\mu\nu} T_{\mu\nu}|_{\Sigma},
\end{equation}
so that the geodesics cross $\Sigma$ when $x = 0$, and $\delta(x)S_{\mu\nu}$. For further details, we refer the
reader to \cite{Bejarano:2016gyv}. The quantity $\Theta(x)$ is known as the Heaviside step function whereas $S_{\mu\nu}$ is the surface stress-energy tensor on the thin-shell.

It is interesting that this way the curvature of spacetime becomes divergent at $\Sigma$ for thin-shell wormholes (because the Riemann tensor is singular). But this divergence is physically interpreted as a surface layer with a stress-energy tensor $T_{\mu\nu}|_{\Sigma}$ on it.  Therefore, the existence of curvature divergences exists at the wormhole throat. 

\begin{figure}[t]
\includegraphics[width=7.5cm]{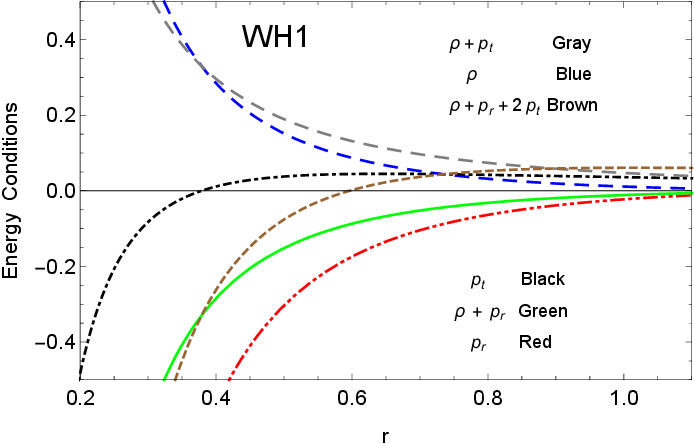}
\caption{Variation all the physical quantities and energy conditions with $A=1.23$, $\chi =-2,~ \omega =-2,~c_3=7.74$ for WH1.}\label{fig5}
\end{figure}

\begin{figure}[t]
\includegraphics[width=7.5cm]{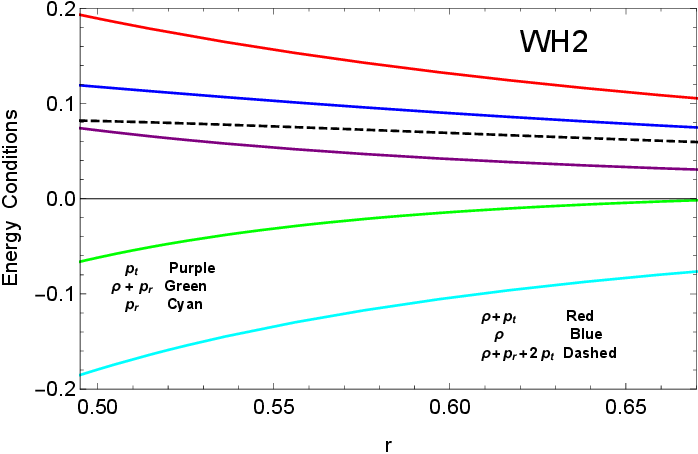}
\caption{Variation all the physical quantities and energy conditions with $B = -0.44,~\chi =-2,~ n =-0.4,~c_3=-10$ for WH2.}\label{fig6}
\end{figure}

\section{Conformal Symmetry Wormhole}\label{sec6}
In general, to solve the three field equations Eqs. (\ref{eq24}-\ref{eq26}) with the following four unknown functions of $r$, namely, $\rho$, $p_r(r)$, $p_t(r)$ and $\psi$ is mathematically well-defined problem. For obtaining an explicit solution one has to specify or determine the EoS, the shape function $b(r)$ \textit{etc.} by implementing some physical
conditions. We employ the following approach to extract and analyze the solutions as below. 

\subsection{On spherical wormhole with isotropic pressure}

The case of a isotropic wormhole i.e. when $p_r = p_t$ is particularly simple one, yet it provides enough interesting results \cite{Cataldo:2016dxq}. In order to analyze solutions 
we shall now on take into consideration  Eqs. (\ref{eq24}) and (\ref{eq25}), which yield
\begin{equation}
\psi^2=\frac{C_{3}^{2}}{2}\left[1-\left(\frac{r}{r_0}\right)^2\right] ,
\end{equation}
where the constant term is determined by imposing the condition $\psi(r_0)=0$ at the throat of the wormhole. Now, using the condition  in Eq. (\ref{eq19}),
we obtain the form of shape function as
\begin{equation}\label{eq31}
\frac{b(r)}{r}=\frac{1}{2}\left[1+\left(\frac{r}{r_0}\right)^2\right] .
\end{equation}
The aim of this section is to see the behavioral effects of  $b(r)$ and its derivative $b'(r)$. Here the throat of wormhole is located at $r_0$. From the obtained shape function (\ref{eq31}), one can easily check that $b'(r_0) =2 \nless 1$. 

In principle, flaring-out condition at the throat should obey the following inequality $b'(r_0)< 1$, which is not reflecting for isotropic pressure wormhole solution.

\subsection{Wormhole solutions with specific choices}
In the following analysis, we consider the relationship involving specific form of equation of state and anisotropy to solve the field equations.

\subsubsection{WH1: Model with $p_r = \omega \rho$ }

With the definitions of $\rho_{eff}, ~p_{eff}$ and $\mathcal{P}_{eff}$, one can rewrite the field equations (\ref{eq24})-(\ref{eq26}) further in the following form:
\begin{eqnarray}
\rho(r) &=& \frac{2 C_3^2 (\chi +2 \pi )-\psi [r (3 \chi +8 \pi ) \psi'+4 \pi  \psi]}{4 C_3^2 r^2 (\chi +2 \pi ) (\chi +4 \pi )}, \label{rho} \\
p_r(r) &=& \frac{\psi [4 (\chi +3 \pi ) \psi-r \chi  \psi ']-2 C_3^2 (\chi +2 \pi )}{4 C_3^2 r^2 (\chi +2 \pi ) (\chi +4 \pi )}, \label{pr}\\
p_t(r) &=& \frac{\psi [r (3 \chi +8 \pi ) \psi '+4 \pi  \psi]}{4 C_3^2 r^2 (\chi +2 \pi ) (\chi +4 \pi )}. \label{pt}
\end{eqnarray}

Let us start for searching an exact wormhole model by considering a linear EoS which is characterized by $p_r=\omega \rho$. Now, if we take account of (\ref{rho}) and  (\ref{pr}) then,  after integration, we can recover the functional form of $\psi(r)$, which yield
\begin{eqnarray}
\psi(r) &=& \frac{1}{\sqrt{2 \chi +2 \pi  (\omega +3)}} \bigg\{\exp \Big[4 \{\chi +\pi  (\omega +3)\}  \Big(A+ \nonumber \\
&& \frac{2 \log r}{\chi-3 \chi  \omega -8 \pi  \omega }\Big)\Big]+C_3^2 (\chi +2 \pi ) (\omega +1) \bigg\}^{1 \over 2},\label{sol1}
\end{eqnarray}
and the corresponding shape function takes the form
\begin{eqnarray}
{b(r) \over r} &=& 1-{\psi^2 \over C_3^2} = 1- \frac{C_3^{-2}}{2 \chi +2 \pi  (\omega +3)} \times \nonumber \\
&&  \bigg\{\exp \Big[4 \{\chi +\pi  (\omega +3)\} \Big(A+\frac{2 \log r}{\chi-3 \chi  \omega -8 \pi  \omega }\Big)\Big]  \nonumber \\
&&  +C_3^2 (\chi +2 \pi ) (\omega +1) \bigg\}. \label{shap1}
\end{eqnarray}
In this case we have for $r \geq r_0$ the metric component $g^{-1}_{rr} \geq 0$ if $\omega <-1$.  In Fig. (\ref{fig1}), we show the behavior of shape function for $\omega = -2$. This result shows that a wormhole solution requires a phantom-energy background, i.e. $\omega < -1$. The use of phantom-energy is not new in wormhole physics (see refs. \cite{Zaslavskii:2005fs,Cataldo:2008ku,Jamil:2008wu,Lobo:2012qq,Nandi:2016ccg}). The energy density in cosmology setting related to the phantom energy is considered positive, $\rho>0$, and we shall maintain this condition. 

The graphical behavior of the $b(r)-r$, $b(r)/r$, and $b'(r)$ are depicted in Figs. \ref{fig2}-\ref{fig4} for WH1 and WH2. From Fig. \ref{fig3}, we find that $b(r)-r$ cuts the $r$-axis, with the throat at $r_0$ = 0.757 and 5.1, respectively. We also observe that $b'(r) <1 $, which obeys the flaring out condition appear in Fig. \ref{fig4}.  Moreover, we can see directly from Fig. \ref{fig2} that the asymptotic behavior $b(r)/r \rightarrow 0$ as $r \rightarrow \infty$, but the redshift function does not approach zero as $r \rightarrow \infty$, which is expected for conformally symmetric wormhole \cite{Rahaman:2014dpa,Bhar:2016vdn}. This means the wormhole spacetime is not asymptotically flat, so one needs to match these interior
geometries to an exterior vacuum spacetime, at a junction interface which we have discussed in sec \ref{sec5}. 

Thus, in this case the stress-energy tensor components are given by
\begin{eqnarray}
\rho(r) &=& \frac{1}{2 r^2 [\chi +\pi  (\omega +3)]} \bigg\{\frac{3 C_3^{-2}}{\chi  (3 \omega -1)+8 \pi  \omega}+\frac{\chi +2 \pi }{\chi +4 \pi } \nonumber \\
&& \hspace{-0.7 cm} \exp \left[4 [\chi +\pi  (\omega +3)] \left(A+\frac{2 \log r}{\chi-3 \chi  \omega -8 \pi  \omega }\right)\right] \bigg\}, \label{rh1}\\
p_r(r) &=& \omega \rho(r), \\
p_t(r) &=& \frac{r^{-2}}{2  [\chi +\pi  (\omega +3)]} \bigg\{\frac{\pi  (\omega +1)}{\chi +4 \pi }-\frac{3C_3^{-2}}{\chi  (3 \omega -1)+8 \pi  \omega}  \nonumber \\
&& \hspace{-0.8 cm} \exp \left[4 (\chi +\pi  (\omega +3)) \left(A+\frac{2 \log r}{\chi-3 \chi  \omega -8 \pi  \omega }\right)\right] \bigg\} .
\end{eqnarray}
To see in a more quantitative way we also analyzed the energy conditions. 
In Fig. \ref{fig5}, we present the graphical behavior of the NEC, WEC and the SEC in terms of the $\rho$, $P_r$ and $P_t$, for different values of parameters $A=1.23$, $\chi =-2,~ \omega =-2$ and $c_3=7.74$. Fig. \ref{fig5}, shows the validity of $\rho \geq 0$ (blue). With the above solution we also found that $\rho+p_r <0$ but $\rho+p_t >0$ that ensure the violation of NEC and this lead to the violation of WEC also. One can  see from figure that the SEC (brown) is also violated. 

Now, we can construct embedding diagrams to represent a wormhole and extract some useful information for the obtained shape function, $b(r)$.

Considering a fixed moment of time, $t = \textit{const}$ \& $\theta=\pi/2$ and embed the metric into three-dimensional Euclidean space, we obtain the embedding surface which is given by
\begin{eqnarray}
{dz(r) \over dr} = \pm {1 \over \sqrt{r/b(r)-1}}=\pm  {b(r)/r \over \sqrt{1-b(r)/r}}. \label{zr}
\end{eqnarray}
For this particular case, the above equation becomes
\begin{eqnarray}
z(r) &=& \pm \int {C_3 \over \psi} \left(1-{\psi^2 \over C_3^2}\right)~dr, \label{zz}
\end{eqnarray}

\begin{figure}[t]
\includegraphics[width=7.5cm]{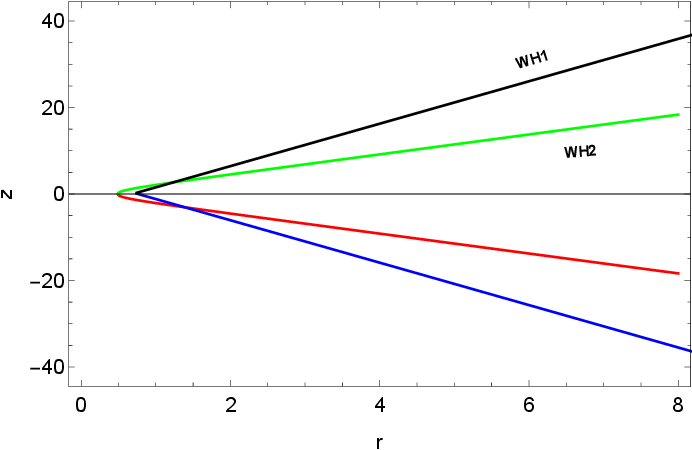}
\caption{Embedding diagrams of two wormholes WH1 and WH2.}\label{fig7}
\end{figure}
\begin{figure}
\centering
\includegraphics[width=0.28\textwidth,height=7cm]{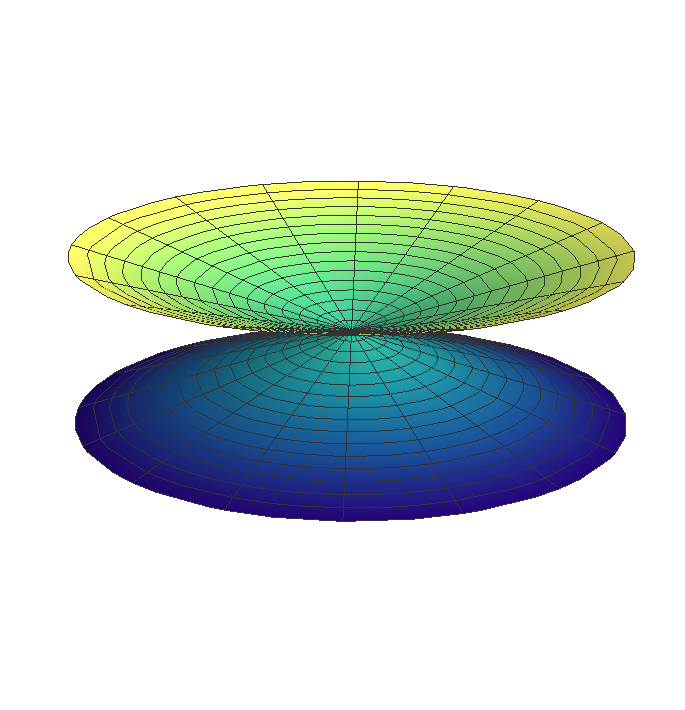}\\
\includegraphics[width=0.3\textwidth,height=6cm]{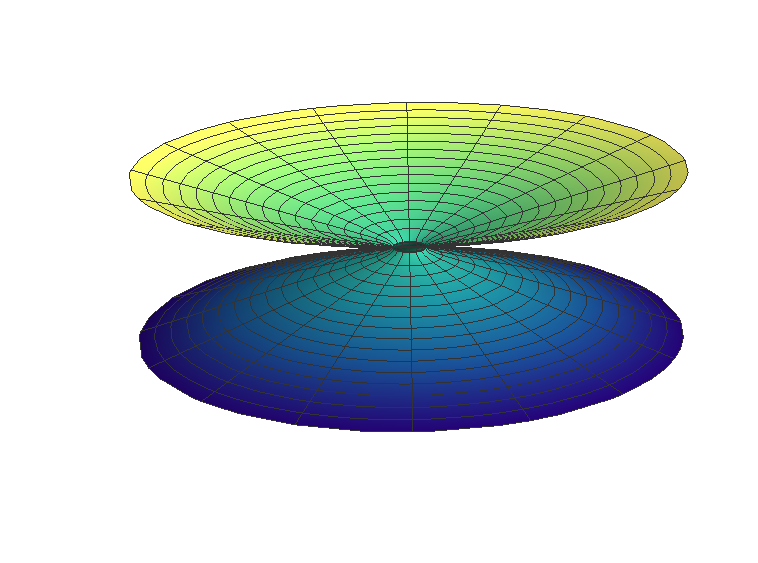}\\
\caption{Embedding surfaces of the two wormholes (WH1 and WH2) after revolution about Z-axis.}
\label{fig8}
\end{figure}


which on integration, we get
\begin{eqnarray}
z(r) &=& \frac{r}{C_3 \sqrt{2 \zeta } [\chi  (3 \omega -5)+4 \pi  (\omega -3)] \sqrt{e^{4 A \zeta } r^{-\sigma }+\eta }} \nonumber \\
&& \bigg[\zeta  \big\{\chi  (5 \omega -3)+8 \pi  (\omega -1)\big\} \sqrt{\frac{e^{4 A \zeta } r^{-\sigma }}{\eta }+1} \nonumber \\
&& 2 C_3^2  \, _2F_1\left(\frac{1}{2},\frac{p}{8 \zeta };\frac{q}{8 \zeta };-\frac{e^{4 A \zeta } r^{-\sigma }}{\eta }\right) \nonumber \\
&& -\big\{\chi  (3 \omega -1)+8 \pi  \omega \big\}\left\{e^{4 A \zeta } r^{-\sigma }+\eta \right\} \bigg],
\end{eqnarray}
where
\begin{eqnarray}
\zeta &=& \chi +\pi  (\omega +3) ~;~ \sigma =\frac{8 \zeta }{\chi  (3 \omega -1)+8 \pi \omega } \nonumber \\
\eta &=& C_3^2 (\chi +2 \pi ) (\omega +1) ~;~p=-3 \chi  \omega +\chi -8 \pi  \omega \nonumber \\
q &=& 3 [8 \pi -\chi  (\omega -3)]. \nonumber
\end{eqnarray}
The embedded surface and surface of the revolution for $z(r)$ about the $Z-$axis are shown in Figs. \ref{fig7} and \ref{fig8}.

\subsubsection{WH2: Model with  $p_t= n p_r $}

Here, we investigate the wormhole solution for a particularly interesting anisotropy, already explored in \cite{Rahaman:2006xa,Moraes:2017mir}, given by
\begin{equation}
  p_t=n p_r,  
\end{equation}
where the state parameter $n$ is a constant. With this assumption and solving the differential equations  (\ref{rho})-(\ref{pt}), as the same procedure for WH1, the function $\psi(r)$ takes the form 
\begin{eqnarray}
\psi(r) &=& \frac{1}{\sqrt{2 n \chi +\pi  (6 n-2)}} \bigg[e^{4 B \Omega } r^{\Lambda } ((n+3) \chi +8 \pi )^{\Lambda }+ \nonumber \\
&& C_3^2 n (\chi +2 \pi ) \bigg]^{1/2},
\end{eqnarray}
where $\Lambda =\frac{8 (n \chi +\pi  (3 n-1))}{(n+3) \chi +8 \pi }$ and $\Omega =n \chi +\pi  (3 n-1)$.\\

Using the definition of $b(r)$ provided in Eq. (\ref{eq19}), one can find the shape function $b(r)$ as
\begin{eqnarray}\label{shap2}
{b(r) \over r} &=& 1-\frac{e^{4 B \Omega } r^{\Lambda } [(n+3) \chi +8 \pi ]^{\Lambda }+C_3^2 n (\chi +2 \pi )}{c_3^2 [2 n \chi +\pi  (6 n-2)]}. \nonumber \\
\end{eqnarray}
As we can see from Fig. \ref{fig2}, that the solutions are asymptotically flat, i.e.
$b(r)/r \rightarrow 0$ as $r \rightarrow \infty$, because of decreasing graphs with increasing 
$r$.  In addition, we plot in Figs. \ref{fig3} and \ref{fig4}, the characteristic picture of the shape function. The red curve represents a regular wormhole solution which cuts $r$-axis at 0.494 is the \textit{throat} of the WH2. As seen in the figure \ref{fig4},  that $b'(r) <1 $, which obeys the flaring out condition. Clearly, in this case also for $r \rightarrow \infty$, the redshift function does not approach zero. Thus, one needs to match this solution to an exterior spacetime at a junction interface, $a > 2M$.

Now, the stress-energy tensor components for the EoS
are given by
\begin{eqnarray}
\rho(r) &=& \frac{1}{2 C_3^2 r^2 (\chi +4 \pi ) \Omega } \Big\{C_3^2 [n \chi +\pi  (2 n-1)]-3 n \nonumber \\
&& (\chi +4 \pi ) e^{4 B \Omega } r^{\Lambda } [(n+3) \chi +8 \pi ]^{\Lambda -1} \Big\}, \\
p_r(r) &=& \frac{3 (\chi +4 \pi ) e^{4 B \Omega } r^{\Lambda } [(n+3) \chi +8 \pi]^{\Lambda -1}+\pi  C_3^2}{2 C_3^2 r^2 (\chi +4 \pi ) \Omega }, \nonumber \\
\\
p_t &=& n p_r.
\end{eqnarray} 
To determine the energy conditions we have plotted graphs, and 
Fig. \ref{fig6} illustrates the behaviour of the null, weak and strong energy conditions. Clearly, in this case we have $\rho > 0$ (blue curve). We are mostly interested in the NEC, because its violation implies the violation of WEC also. In Fig. \ref{fig6}, $\rho+p_r <0$ but $\rho+p_t >0$ i.e. violation of NEC and consequently the WEC, are violated. Interestingly we note that SEC (dashed curve) is satisfied in this case. All solutions are characterized by considering parameter values $B = -0.44,~\chi =-2,~ n =-0.4$ and $c_3=-10$ for WH2.

To further interpret these results let us bring out attention on the embedded surface, which is determined from (\ref{zr}) and found as
\begin{eqnarray}
z(r) &=& \frac{r \sqrt{e^{4 B \Omega } r^{\Lambda } [(n+3) \chi +8 \pi ]^{\Lambda }+C_3^2 n (\chi +2 \pi )}}{C_3 \sqrt{2 n \chi +\pi  (6 n-2)} [(5 n+3) \chi +4 \pi  (3 n+1)]} \nonumber \\
&& \bigg[-e^{4 B \Omega } r^{\Lambda } [(n+3) \chi +8 \pi ]^{\Sigma }+ \bigg\{C_3^2 \bigg(-n (\chi +2 \pi )  \nonumber \\
&& [(n+3) \chi +8 \pi]+\frac{2 [n \chi +\pi  (3 n-1)]}{\chi +2 \pi } \times \nonumber \\
&& [3 (n+1) \chi +\pi  (8 n+4)] \, _2F_1(1,\Gamma ;\Theta ;-\Xi )  \nonumber \\
&& \Big\{\frac{e^{4 B \Omega } r^{\Lambda }}{C_3^2 n} [(n+3) \chi +8 \pi ]^{\Lambda }+\chi +2 \pi \Big\}\bigg) \bigg\}\bigg],
\end{eqnarray}
where for notational simplicity we use
\begin{eqnarray}
\Sigma &=& \frac{3 (3 n \chi +8 \pi  n+\chi )}{(n+3) \chi +8 \pi } ~;~\Gamma =\frac{(5 n+3) \chi +4 \pi  (3 n+1)}{8 [n \chi +\pi  (3 n-1)]} \nonumber \\ 
\Theta &=& \frac{3 (3 n \chi +8 \pi  n+\chi )}{8 [n \chi +\pi  (3 n-1)]} ~;~\Xi =\frac{e^{4 B \Omega } r^{\Lambda } [(n+3) \chi +8 \pi ]^{\Lambda }}{C_3^2 n (\chi +2 \pi )},\nonumber
\end{eqnarray}
which is again well-defined. The embedding diagram and its surface revolution about $Z-$axis are shown in Figs. \ref{fig7} and \ref{fig8}.

\section{Volume integral quantifier}\label{sec7}
It is convenient to consider the ``volume integral quantifier" to know how much of exotic matter is required to support a traversable Lorentzian wormhole on a local scale. This was first prompted by Visser \textit{et al} \cite{Visser:2003yf}. Later, a more technical review was proposed in \cite{Nandi:2004ku}.
Quantifying the amount of exotic matter has been considered by 
the following defined integral $I_V =\int\left(\rho(r)+p_r(r)\right)\mathrm{d}V$, and
with a cut-off of the stress-energy at $a$ is given by
\begin{eqnarray}
I_V  &=& \left[r\left(1-\frac{b}{r}\right)\ln\Big(\frac{e^{\nu}}{1-b/r}\Big)  \right]_{r_0}^{a}\\\notag
&-&\int_{r_0}^{a}\left[(1-b')\ln\Big(\frac{e^{\nu}}{1-b/r}\Big)  \right]\mathrm{d}r,
\end{eqnarray}
where $\mathrm{d}V=r^2\sin\theta~ \mathrm{d}r ~\mathrm{d}\theta ~\mathrm{d}\phi$, and the boundary term at $r_0$ vanishes by our construction as $b(r_0)=r_0$. Then, the volume-integral reduce to (see Ref. \cite{Boehmer:2007md} for more details)
\begin{eqnarray}
I_V  &=& \left[a\left(1-\frac{b(a)}{a}\right)\ln\Big(\frac{e^{\nu(a)}}{1-b/a}\Big)\right]\nonumber \\
&& -\int_{r_0}^{a}\left[(1-b')\ln\Big(\frac{e^{\nu}}{1-b/r}\Big)  \right] \mathrm{d}r,  
\end{eqnarray}
Taking into account the redshift function $e^\nu = C_2^2 r^2$, and the form function, Eqs. (\ref{shap1}) and (\ref{shap2}),
we obtain the following expression
\begin{eqnarray}
I_{V}(WH1) &=& \left[a\left(1-\frac{b(a)}{a}\right)\ln\Big(\frac{e^{\nu(a)}}{1-b/a}\Big)\right]-\nonumber \\
&& \bigg[\frac{r (\chi +2 \pi ) (\omega +1)}{\zeta }-\frac{(\chi +4 \pi ) (\omega +1)e^{4 A \zeta }}{C_3^2 \zeta  [8 \pi -\chi  (\omega -3)]} \nonumber \\
&&  r^{q/p}-\frac{r \left[e^{4 A \zeta } r^{-\sigma }+C_3^2 (\chi +2 \pi ) (\omega +1)\right]}{2 C_3^2 \zeta } \nonumber \\
&& \log \left(\frac{2 C_2^2 C_3^2 \zeta  r^2}{e^{4 A \zeta } r^{-\sigma }+C_3^2 (\chi +2 \pi ) (\omega +1)}\right)\bigg]_{r_0}^a \\
I_{V}(WH2) &=& \left[a\left(1-\frac{b(a)}{a}\right)\ln\Big(\frac{e^{\nu(a)}}{1-b/a}\Big)\right]- \nonumber \\
&& \bigg[\bigg\{\frac{2 (n-1) (\chi +4 \pi ) e^{4 B \Omega } r^{\Sigma } }{(3 n \chi +8 \pi  n+\chi)[(n+3) \chi +8 \pi ]^{-\Sigma } }  \nonumber \\
&& -2 C_3^2 n r (\chi +2 \pi ) [(n+3) \chi +8 \pi ]+  \nonumber \\
&& \bigg( e^{4 B \Omega } [(n+3) \chi +8 \pi ]^{\Lambda } r^\tau+C_3^2 n (\chi +2 \pi ) \bigg) \nonumber \\
&& \hspace{-0.8 cm}\log \left[\frac{2 C_2^2 C_3^2 r^2 \{n \chi +\pi  (3 n-1)\}}{e^{4 B \Omega } r^{\Lambda } [(n+3) \chi +8 \pi ]^{\Lambda }+C_3^2 n (\chi +2 \pi )}\right]\bigg\} \nonumber \\
&& \bigg[{2 C_3^2 \{n \chi +\pi  (3 n-1)\} \over \{(n+3) \chi +8 \pi \}^{-1}} \bigg]^{-1}\bigg]_{r_0}^a,
\end{eqnarray}
where $\tau={\frac{8 \Omega }{(n+3) \chi +8 \pi }}$. It is interesting to note that when $a \rightarrow r_0$ then $I_V \rightarrow 0$ for both cases. In fact, one can also observe that for WH1 if the parameter $\omega$ arbitrary close to $-1$, the integral may be infinitesimally small.  These results fundamentally confirm the validity of conformally symmetric phantom wormhole solutions, as described in \cite{Lobo:2005us,Lobo:2005yv},
where the violation of ANEC is arbitrarily small when the interior solution is matched to an exterior vacuum spacetime.

\section{SUMMARY AND DISCUSSION}\label{sec8}

In the present paper, we investigate the possible
existence of wormhole solutions in the framework of $f(R, T)$ gravity under the assumption of spherical symmetry and the existence of a conformal Killing symmetry.
To address the problem we consider a particular
and simple model $f(R,T)=R+2\chi T$, where $R$ is the Ricci scalar and $T= -\rho+p_r+2p_t$  denotes the trace of the energy–momentum tensor of the matter content. Even within this simple theoretical model the field equations become extremely complicated, and therefore conformal symmetry is a more systematic approach in searching for exact analytic solution. The obtained solutions in this article are not asymptotically flat, where
distribution of the exotic matter restricted to the throat neighborhood, and we consider a cut-off of the stress-energy tensor at a junction interface by matching an interior traversable wormhole geometry. In fact, we are successfully able to make the a particular asymptotically flat wormhole geometries where the dimensions are not arbitrarily large.

Next, we explore and analyze two cases separately. At the first part,  the obtained wormhole solutions are constructed for the matter sources with isotropic pressure. However, showing explicitly that the solution violates the basic criteria for wormhole. Further, we proceed by introducing an EoS relating with pressure (radial and lateral) and density. We show the possibility of having traversable wormhole geometries supported by phantom energy. In this case, the energy density  $\rho \geq 0$ is positive which consequently violates the null energy condition. However, we emphasize that when $\omega \rightarrow -1$ the volume integral quantifier would by itself become arbitrarily small i.e. theoretically it is possible to construct these geometries with 
vanishing amounts of ANEC. For our convenience we have also analyzed physical properties and characteristics of traversable wormholes by using graphical representation (see Fig. \ref{fig1}-\ref{fig8}). 

In the second part of the paper we obtain a similar picture for the models described by $p_t= n p_r $. Still in this case, obtained solution are violating the NEC and WEC with the  energy density  $\rho \geq 0$, but interestingly satisfying the SEC. From our analysis it is very transparent that the assumption of a static conformal symmetry, i.e., 
with a static vector $\xi$, is found responsible to find an exact solutions of traversable wormholes.

\subsection*{Acknowledgments}
FR  would like to thank the authorities of the Inter-University Centre for Astronomy and Astrophysics, Pune, India for providing the research facilities.  FR is  also thankful to DST-SERB,  Govt. of India and RUSA 2.0, Jadavpur University,  for financial support.

\end{document}